\documentstyle[11pt,paspconf]{article}

\begin{document}

\begin{center}
\vspace*{-1.9cm}
\begin{footnotesize}\baselineskip 10 pt
Proc.~~{\it Library \& Information Services III}, ~Tenerife, Spain, April 21--24, 1998\\[.5ex]
~~~~~~~to appear in ASP Conference Series~~~~~~(1998),~~~~ \\[.5ex]
eds.\ U.\,Grothkopf, H.\,Andernach, S.\,Stevens-Rayburn, \& M.\,Gomez \\[1ex]
\end{footnotesize}
\end{center}

\vspace*{0.4cm}

\title{Library \& Information Services: an Astronomer's Wishlist}

\author{Heinz Andernach}
\affil{Depto. de Astronom\'\i a, IFUG, Guanajuato, C.P.~36000, Mexico}

\begin{abstract}
I review some of the past and current methods for retrieval of literature
and other published information, excluding commercial services.  Much of
this is a personal view and based on experience made at various
institutions, some of them neither with an adequately complete astronomy
library, nor with a professional astronomy librarian. Rather than describing 
current retrieval methods, a few of their weaknesses are identified 
which merit future improval.  Despite the availability of
powerful electronic tools, we need to improve efforts in safeguarding
published numerical and textual information in a format readily
usable by astronomers.  Comments are made on a user-friendly arrangement 
of a library, and on useful tasks for librarians with available time.  
\end{abstract}


\keywords{libraries, electronic information systems, bibliographies,
published tabular data, electronic preprints, electronic journals}


\section{Diversity of Libraries and Information Systems: 
  A Personal View} \label{diversity}

In this section I describe some of the experience relevant to the subject, 
which I have made at various astronomy institutions over the last 20 years.

At Max-Planck Institut f\"ur Radioastronomie Bonn, Germany (1976--86) 
I became used to both a good library (complemented by the 
Astronomische Institute of University of Bonn, next door) and
an excellent librarian (of the pre-electronic information age).  In the early
1980s, the Fachinformationszentrum (FIZ) Karlsruhe started to offer a
service providing a monthly extract of the latest literature on a
user-specified combination of keywords (a ``user profile''), which at that
time was free of charge for users at MPIfR. However, due to the comprehensive
collection of journals and preprints received at the MPIfR library in the
field of radio astronomy, there was hardly any article listed by FIZ that I
had not seen before. It was mainly the biannual volumes of {\it Astronomy
\& Astrophysics Abstracts} (AAA) which always offered some new references
to me, although with the usual delay of about eight months after the end of
the nominal half-year period covered by each edition of AAA.
During this ``pre-Internet'' age, any exchange of bulk data or information
was much slower and more tedious than today. I recall that the computing center
provided about two major astronomical catalogs (to be interrogated via
batch jobs).  Requesting additional catalogs implied long delays,
exchange of tapes, or the conversion of formats, to be compatible with
local hard- and software.

In 1987/8, at the Inst.~Argentino de Radioastronom\'\i a, facilities were
profoundly different due to economic limitations.  The collection of
journals was limited to the very core journals, which arrived irregularly,
and often with several months of delay.  However, during a visit to
C\'ordoba (Argentina) I enjoyed the astronomy library when working on a
project that required browsing many years of the core journals in
astronomy. Tables and chairs in almost every corridor of the
library allowed one to study the volumes right next to the shelves.

I entered the Internet age (then ``BITnet'') when working at the 
Astronomy Deptartment of Instituto de Pesquisas Espaciais (INPE, Brazil), which 
had a decent collection of astronomical journals but was rather 
poor in preprints. By access to electronic mail, from 1989 on, I came across
the listing of preprints received by the Space Telescope Science Institute (STScI)
and distributed as the ``STEP sheet'' every two weeks by its librarian.
Due to this unique service, I felt again up-to-date in astronomical literature,
and together with the electronic mail directory of astronomers
(the {\it RGO Email Guide}, by C.~Benn \& R.~Martin), it allowed me to obtain 
reprints or preprints from the authors rather efficiently.
It is amazing that until today I manage to
surprise some colleagues when telling them about the STEP sheet of which they 
had not heard before. 
However, for older and rarer literature, I used to visit
the library of Inst.\ Astron.\ e Geof\'\i sico (IAG, S\~ao Paulo) with its 
exceptional collection of journals
and observatory reports from all over the world. It was there that I discovered
the usefulness of the journal {\it Current Contents of Physical, Chemical
\& Earth Sciences} to browse the contents pages of recent issues of virtually 
all relevant journals in astronomy. Email allowed me to get hold of an 
article of interest usually within weeks.
In some cases librarians of large astronomical institutions 
were kind enough to send me copies of articles I did not have access to, 
usually upon my request via email.

Exchange of electronic copies of articles via email was still in its
infancy in 1990 both due to network limitations and format
incompatibilities. When I received the first email containing an article
written in \TeX, I had to carry it on a diskette to a friend who happened
to have \TeX\ installed on his PC.  Within less than a year of its first use,
email had turned indispensable for me to maintain collaborations and be
alerted about the latest preprints. It allowed me to start an international
campaign to safeguard tabular data from published journals in electronic
form, which soon led to the most complete collection of radio source
catalogs in existence (\S\ref{pubdata}), and email led me to find an
opportunity to continue my research at Instituto de Astrof\'\i sica de Canarias,
(IAC, Tenerife, Spain) in 1991. IAC possesses the biggest modern astronomy 
library in Spain, with a large variety of astronomical journals 
including physics, electronics and computer sciences, and volumes of 
recent proceedings in astrophysics.  During my two years at IAC, 
none of my many suggestions for purchasing books were rejected. 
IAC also had a large collection of astronomical newsletters providing 
material for a review article on
``Network Resources for Astronomers'' (Andernach et al.\ 1994).  In the spring
of 1993 the World Wide Web (WWW) had been created by researchers at CERN, and
it was only due to the up-to-date level of software maintained at IAC, that
I was able to access the information (offered {\it only} on WWW) about the
ADASS-III meeting I was going to attend.

In 1992 the Los Alamos National Laboratory (LANL, USA) and the
International School for Advanced Studies (SISSA, Trieste, Italy) started to keep
mirror archives of electronically submitted preprints.  Initially these
were limited to the fields of theoretical cosmology and particle physics,
and were not known to many astronomers (including myself), but have now become
impressively popular for the deposit and circulation of astronomy preprints
(\S\ref{versprob}).

At the Observatoire de Lyon (1993/94), I enjoyed its excellent collection of
astronomical journals, many of them available back to the very first volume.  
What I did not recall from the other libraries was the full collection of 
{\it Astronomischer Jahresbericht}, the predecessor of AAA from 1899 through
1968. Despite this I recall the disadvantage that the main room with
journals and textbooks had no table or chair where one could study (and not
even space for it) and the next decent copy machine was two floors up. 
Observatoire de Lyon also had a good viewing facility for the Palomar and 
ESO/SERC Optical Sky Survey. Much of this can now be done with the {\it Digitized
Sky Survey} on WWW, but the original survey plates, prints or films still
offer a finer detail e.g.\ for morphological classification of galaxies and
should by all means be preserved.

In 1995/96, working at the IUE Observatory in Spain, I saw the advent of
electronic journals like {\it ApJ Letters}, soon followed by the 
{\it New Astronomy} and the {\it ApJ}. I was surprised to find that 
some large data tables in MNRAS were still published on {\it microfiche}, in
fact as late as early 1997, at a time when microfiche copiers had
practically disappeared from astronomy libraries.  When trying to get
hardcopies of several data sets published on microfiche over the last 20
years I found that STScI was among the few institutions that still had such
a machine available! For how much longer will these machines exist\,?

Since 1996, I have worked in a small but growing Astronomy Department at
Guanajuato University, Mexico. Although various donations of sets of journals
had been collected by the founders of the Department before my arrival,
I was surprised that I was the first person to show an active interest in
putting all these donations together, as soon as we found space for them. 
Apparently, NASA's Astrophysics Data System (ADS) article service 
could already satisfy most of the needs
of our department members, and it required a person of ``special'' interests
to go after the needs of the library. While this may show a (perhaps temporal?)
tendency (or even a need) for modern researchers to survive without a
library, I continue my contacts to obtain further donations to complete our
physical library, and wish to take the opportunity to thank several members of
the audience who have donated material to our department. 
In my current position, I am also responsible for the library
for the first time, and admit that I do not fulfil many of the 
requirements proposed further below. 

Details and URLs on many more information resources in astronomy can be found
in Andernach (1998). In the URLs quoted here I omit the~ http://\,.

\section{The Physical Library}

The following list of requirements for the functionality of a physical library 
may seem all too obvious, but as I do not often find them,
I mention them anyway.

The library should be a {\it study room} rather than a {\it storage room},
and should invite one to look at more than just that piece of literature one is
searching for.  Tables and chairs should be present in various corners to
allow one to read and work next to the material found. Obviously a copy machine
should be reasonably close.  The item searched for should be {\it easily
locatable} at any time, and without the presence of the librarian.  Ideally,
the library catalog should be searchable in electronic form, accessible 
from any user's desktop (including one in the library itself), e.g.\ via
the WWW.  The search program should be self-explanatory, and the code for 
the physical location of a certain book or item should be clearly marked on the
shelves.  The current location of books on loan should be made known to the
interested user, either by replacing the book with a card containing the
name of the user who lent the book, or by indicating this user in the
catalog record.  A publicly accessible ASCII file with the library 
catalog, if not too large, is much better than nothing.  Access to the
library should be possible at all times, including nights and weekends,
perhaps via a key or access code outside normal working hours. The reaction
to books disappearing should NOT be access restriction, but ``razzias''
through the staff offices, and strict measures should be applied to users
who do not adhere to the rules.

For very literature-demanding research a computer terminal in the library 
would allow users to interact directly with their files.

In addition to the shelf with the most recent issues of journals, the recently
acquired books should also be displayed separately for at least a couple of
months. Alternatively a listing of new acquisitions could be distributed to
the users on a regular basis.  Newsletters and other ``grey'' literature
should be archived and made available to users if possible. Duplicate items
should be kept and a list be posted to networks of librarians.

\section{Additional Services the Librarian could provide} \label{addserv}

Over the last decade about a handful of astronomy librarians have provided 
services of relevance to the entire astronomical community. Examples
are the STEP sheet 
(sesame.stsci.edu/lib/stsci-preprint-db.html), 
the Astronomy Thesaurus \linebreak[4] 
(msowww.anu.edu.au/library/thesaurus/),
the list of astronomy meetings \linebreak[4]
(cadcwww.dao.nrc.ca/meetings/meetings\_without.html),
a data\-base of book reviews 
(www.astro.utoronto.ca/reviews1.html), 
and the directory of astronomy libraries
(www.eso.org/libraries/astro-addresses.html).
These very successful, though time-consuming projects will hopefully
stimulate further librarians to take up projects of use to
a larger community. To avoid duplication, these activities should be
coordinated with the established networks of librarians.
Apart from such outstanding efforts like the above,
there are services librarians could provide to the local users, 
if their time allowed.  I describe a few here.

Not many libraries can afford to subscribe to virtually all astronomy
journals. The exchange of paper preprints seems to be decreasing in
favour of electronic circulation. This means that a large fraction of
current astronomical literature has to be searched for at other places.  The
contents pages of an increasing number of astronomy journals are accessible
in one way or another, and could be displayed in the library together with
the other journals.  A more expensive method is to subscribe to 
{\it Current Contents} which offers the contents pages of practically all
astronomy journals.  An alternative solution is to regularly consult the
URLs of journals offering their contents pages on WWW and prepare
printouts. The {\tt UnCover} database (uncweb.carl.org)
includes most astronomy-related journals and allows one to browse their
contents by issue. Listings of preprints submitted to the LANL/SISSA
preprint server (xxx.lanl.gov) could also be displayed at a suitable
frequency (bi-weekly or monthly). Preprint lists compiled by the librarians
of NRAO and STScI (the ``RAP'' and ``STEP'' sheets) would complete the
collection.  While many researchers are already making use of the latter
facilities individually, the librarian could motivate other astronomers who
are either unfamiliar with these services or pretend to have no time to
consult them.  Since the librarian is frequently responsible for the
publication list of an institute, these publications could be listed and
offered electronically from the library web page. To guarantee widespread
notification of these preprints authors would ideally deliver their
preprints to the LANL/SISSA server and the library page would simply have a
pointer to that location.

The librarian should be accessible by electronic mail and make regular use
of it, e.g.\ to subscribe to relevant distribution lists (Astrolib, EGAL, PAMnet),
to alert colleague librarians about missing items, or to organize the exchange of
duplicate material between libraries.  Equally important for librarians
will be the regular study of relevant information available on the
Internet, e.g.\ on journals, publishers, subscriptions, societies, other
libraries, etc.

\section{Electronic Journals and Preprints}

The appearance of more and more journals in electronic form has stimulated
an intense discussion about the problem of how to safeguard their content
for future generations (e.g.\ Grothkopf 1997).
I shall not enter this discussion here but only mention a few practical 
problems I encountered while using them.

\subsection{The Version Problem for Preprints} \label{versprob}

The problem of different versions of papers being circulated has been with
us since the age of paper-only preprints.  Front-line researchers are
obliged to take much of their latest knowledge from preprints received from
colleagues or found in the library.  Most likely any changes between
preprint and the actually published version will remain unnoticed by the
researcher quoting the original preprint.  In some cases, the part of the
text referred to may have been altered or even have disappeared, and in
rarer cases the referred paper may never be published because the preprint
submitted was never accepted.  With the advent of the LANL/SISSA server of
preprints this situation has not changed.  At least it is now possible to
refer to a unique reference on the LANL server, and to retrieve a paper for
the foreseeable future.  In fact, the reference section of published papers
contain an increasing number of pointers to astro-ph/{YY}{MM}{NNN}.  
The authors may change or even withdraw a
submitted paper, and future retrievals of the paper will report about this
history. However, there are quite a few preprints on these servers, for
which the authors do  not state where the paper is supposed to appear and
whether the deposited version is a submitted or an accepted one.  In
December 1997, the journal {\it New Astronomy} suggested that if authors
wish to deposit their papers for this journal at an electronic preprint
server prior to publication, they should {\it not} deposit the {\it
authentic} accepted version.  Clearly, this is not the kind of
``generosity'' astronomers look for, and it unnecessarily increases the
confusion between preprint and published versions.

\subsection{Ready Access to Published Data} \label{pubdata}

The importance of a central archive for astronomical data of general
interest like object catalogs, survey images, spectra or time series had
led to the creation of data centers like CDS and NASA-ADC in the early
1970s.  With the subsequent development of databases like SIMBAD, NED,
LEDA, etc., there was a growing need for electronic versions of all kinds
of data sets in the published literature, in order to attach these data (or at
least their bibliographic reference) to the respective object, so that a
query on this object would return the article or even the measurement made
on it.  Thus the collection activity of data centers has gradually been
expanded from larger general-purpose catalogs to smaller and more
specialised tables in the published literature. Unfortunately during the
past decade, many chances have been lost to archive both data tables and
full-text versions of articles in electronic form.  These obviously existed
for the large majority of publications since the mid-1980s, but were not
archived for future use.  Only with the advent of the truely electronic
editions some limited full-text search engines have now become available,
like that for the {\it Electronic ApJ} (www.journals.uchicago.edu/ApJ).  
It is reassuring to see that the ADS article service 
(adsabs.harvard.edu/article\_service.html) is approaching its goal
of providing free access to scanned images of all major astronomy journals
back to their volume 1. However, the conversion of these images to
searchable text via {\it Optical Character Recognition} software (OCR)
appears unrealistic.

The recovery of the data sections of articles not previously available at
data centers in electronic form appeared more feasible and urgent to me,
and since 1990 I have collected such data tables and catalogs in the area
of extragalactic objects and radio sources. The radio data constitute the
basis for the presently most complete Internet service for radio sources
(``CATS''; cats.sao.ru). In the early 1990s, I received
most tables upon request to the individual authors, while more recently an
increasing fraction of such data sets may be obtained from the LANL/SISSA
preprint server (risking that these versions may not be identical to the
published ones). A substantial fraction of these tables, usually older ones,
was recovered via a scanner and OCR, followed by a strict proof-reading of
the result before being released to other users, databases or data centers.
Many  tables were printed in such a small font or poor quality that even
recovery with OCR is impossible.  Currently the collection contains data
sets from nearly 900 articles, of which only $\sim$25\% are also
available from established data centers such as CDS and NASA-ADC. About 15--20
data sets are being added per month, and my activity has turned into that
of an independent data center, proving that the capacity of established data
centers is not sufficient to cope with the number of data sets published.
Not surprisingly, the effort to obtain these data sets from the authors is
being duplicated by managers of databases mentioned above and by other
users interested in individual data sets. A better coordination between
data centers, publishers and authors/users is badly needed.

A good example has been set in 1994 with the agreement between the
publishers of A\&A and A\&AS and CDS to archive data tables published in
these journals at CDS. During the last four years, a similar procedure was
followed by the journals ApJ, ApJS, AJ and PASP who published larger data
sets separately on CD-ROM, although at a low level of completeness and with
a delay implied by the CDs being published only twice a year. Since 1998,
the publication of these CDs has been abandoned and the user is supposed to
download the data from the electronic versions of these journals, carrying
with it two drawbacks: users must subscribe to the electronic version,
and the downloaded data are not always in a format directly usable. A few
tests I made at institutions subscribed to electronic journals 
were not too promising.  Downloading a table from the electronic
ApJ as ``text'' results in a file with entries separated by blank lines,
with columns not well justified and some HTML symbols left over. The AJ
actually offers hyperlinks to ASCII versions of its tables, but I found
them to contain TAB symbols requiring some editing to align the columns.
For papers whose data tables were published on the AAS CD-ROM series, the
electronic article showed only the sample page of the table (as the printed
version) and did not offer a link to the full files as on CD-ROM, which is
an unnecessary limitation given the small file sizes involved.  The
electronic A\&A and A\&AS offer links to the CDS electronic archive for the
larger tables, usually those in A\&AS. However, smaller tables mixed within
the text are often inserted as images and papers are not retrievable in
ASCII format at all. MNRAS presently offers only a few sample articles in
electronic form and has not foreseen any archiving means for tabular data
it has published.

While electronic journals have made it possible to {\it view\,} articles much
more timely than previously, they can still be improved to allow 
{\it working} with the data they contain. After having converted hundreds 
of tables from \TeX\ format to ASCII, I wonder whether we really need this
laborious chain of conversions which starts with the marking up of the
original author's ASCII table into \TeX\ for the preprint, and is followed by
a publication in HTML. Later, an interested user downloads the paper in PDF
or PS format and eventually needs a document converter (I am not aware of
reliable ones) to recover the original ASCII table. Clearly, the journal
publishers are mainly interested in selling their journals and will not be
inclined to collaborate in this effort. A possible solution may be to
oblige the authors, as a condition for publication, to provide their data
to the established data centers, including a comprehensive documentation.

\section{Bibliographical Services} \label{bibserv}

Until about 1993, browsing the AAA was about the only means for
bibliographic searches ``without charge'' (except for the cost of the
books). In 1993 the ADS abstract service (\S\ref{diversity})
became accessible via Internet inside USA and was
released on WWW for free access world-wide early in 1994. It is by far
the most popular bibliographic service in astronomy (Eichhorn 1997).
The latter is mainly due to its powerful search tools through well over
a million abstracts in astronomy, instrumentation, physics and geophysics,
covering large parts of the literature back to the mid-1970s.

However, when one needs ``grey literature'', proceedings, etc., or
one wants to measure the productivity of researchers, either from their 
publication records or from citations to their papers, the
ADS, and even commercial services become incomplete, especially for
proceedings papers.

In reaction to the recent threat that AAA may stop publication in 1999 \linebreak[4]
(www.eso.org/libraries/iau97/libreport.html),
some astronomy librarians performed a comparison study 
and found that, in particular, information about conference proceedings 
and observatory reports
found in AAA is missing from ADS and even from the commercial service INSPEC.
For many years, astronomers have been waiting for the AAA to become 
available in electronic form.
Only a few weeks before LISA~III the Astronomisches Rechen-Institut (ARI)
announced ARIBIB, a database of references from AAA since 1983 
(Demleitner et al., these proceedings).
ARI plans to prepare in electronic form the entire content of AAA, 
back to 1899 from {\it Astronomischer Jahresbericht}. 
This will be an invaluable tool at least for those
astronomers working at institutions subscribed to the printed AAA. 
Hopefully these privileges can be levelled out in future
via funding organizations, perhaps in a similar way as has gradually
been achieved in the past to provide free access to SIMBAD.

Drawbacks of most bibliograhical services are their incompleteness 
and large delays in the inclusion of conference proceedings.
The ADS now makes an effort to include references to conference papers 
even before they go into print. It is desirable that such papers 
may be retrieved by the names of the editors or the conference title, 
and be linked to the full table of contents of the volume.

\section{Conclusions}

I have tried to identify some areas in astronomy information systems and
archiving methods that merit future improvement. In astronomy, the evolution
of these systems trace a delicate path between economic interests of
publishers and activities of individual researchers, librarians, or funding
organizations to create non-profit systems for the astronomy community.
The voluntary effort of a handful of astronomy librarians has provided
important contributions to general information services, stimulated by
efficient and instantaneous means of communication (Email) and
``publication'' (WWW). I hope that conferences like the present one will
stimulate further concerted activities of that sort.  It is important to
maintain an active communication between publishers, astronomers, and
librarians to guarantee that new products like electronic journals continue
to serve their consumers well.

\acknowledgments

I am indebted to the meeting organizers for a travel grant, and I would like
to thank A.\,P.~Fairall for his careful reading of the text.

\end{document}